\documentclass[twocolumn,english,superscriptaddress]{revtex4-2}
\usepackage[T1]{fontenc}
\usepackage{amsfonts}
\usepackage{amsmath}
\usepackage[latin9]{inputenc}
\usepackage{textcomp}
\usepackage{amstext}
\usepackage{subscript}
\usepackage{graphics}
\usepackage{graphicx}
\usepackage{float}
\usepackage{color}
\usepackage{pifont}
\usepackage{pmboxdraw}
\usepackage{amssymb}
\usepackage{esint}
\usepackage{tabularx}
\usepackage[export]{adjustbox}
\usepackage{hyperref}
\usepackage{upgreek,textgreek}
\usepackage{siunitx}
\usepackage{multirow}
\usepackage[section]{placeins}

\makeatletter

\newcommand{\lyxmathsym}[1]{\ifmmode\begingroup\def\b@ld{bold}
  \text{\ifx\math@version\b@ld\bfseries\fi#1}\endgroup\else#1\fi}

\usepackage{hyperref}
\usepackage{upgreek,textgreek}
\hypersetup{colorlinks=true,linkcolor=blue,citecolor=blue,urlcolor=blue}
\renewcommand{\fnum@figure}{FIG.~\thefigure}
\makeatother

\usepackage{babel}
\begin{document}
\title{Charge fluctuations above $T_\mathrm{CDW}$ revealed by glasslike thermal transport in kagome metals $A$V$_\mathrm{3}$Sb$_\mathrm{5}$ ($A$ = K, Rb, Cs)}
\author{Kunya Yang}
\thanks {These authors contributed equally to this work.}
\affiliation{Low Temperature Physics Lab, College of Physics \& Center of Quantum
Materials and Devices, Chongqing University, Chongqing 401331, China}

\author{Wei Xia \textcolor{blue}{\textsuperscript{*}}}
\affiliation{School of Physical Science and Technology, ShanghaiTech University, Shanghai 201210, China}
\affiliation{
 ShanghaiTech Laboratory for Topological Physics, Shanghai 201210, China}

\author{Xinrun Mi}
\affiliation{Low Temperature Physics Lab, College of Physics \& Center of Quantum
Materials and Devices, Chongqing University, Chongqing 401331, China}

\author{Long Zhang}
\affiliation{Low Temperature Physics Lab, College of Physics \& Center of Quantum
Materials and Devices, Chongqing University, Chongqing 401331, China}

\author{Yuhan Gan}
\affiliation{Low Temperature Physics Lab, College of Physics \& Center of Quantum
Materials and Devices, Chongqing University, Chongqing 401331, China}

\author{Aifeng Wang}
\affiliation{Low Temperature Physics Lab, College of Physics \& Center of Quantum
Materials and Devices, Chongqing University, Chongqing 401331, China}
\author{Yisheng Chai}
\affiliation{Low Temperature Physics Lab, College of Physics \& Center of Quantum
Materials and Devices, Chongqing University, Chongqing 401331, China}

\author{Xiaoyuan Zhou}
\affiliation{Low Temperature Physics Lab, College of Physics \& Center of Quantum
Materials and Devices, Chongqing University, Chongqing 401331, China}

\author{Xiaolong Yang}
\email{yangxl@cqu.edu.cn}
\affiliation{Low Temperature Physics Lab, College of Physics \& Center of Quantum
Materials and Devices, Chongqing University, Chongqing 401331, China}

\author{Yanfeng Guo}
\email{guoyf@shanghaitech.edu.cn}
\affiliation{School of Physical Science and Technology, ShanghaiTech University, Shanghai 201210, China}

\author{Mingquan He}
\email{mingquan.he@cqu.edu.cn}
\affiliation{Low Temperature Physics Lab, College of Physics \& Center of Quantum
Materials and Devices, Chongqing University, Chongqing 401331, China}
\date{\today}

\begin{abstract}
We present heat capacity, electrical and thermal transport  measurements of kagome metals $A$V$_\mathrm{3}$Sb$_\mathrm{5}$ ($A$ = K, Rb, Cs). In the all three compounds, development of short-range charge fluctuations above the charge density wave (CDW) transition temperature $T_\mathrm{CDW}$ strongly scatters phonons via the electron-phonon coupling, leading to the glasslike phonon heat transport, i.e., the phonon thermal conductivity decreases weakly upon cooling. Once the long-range charge order sets in below $T_\mathrm{CDW}$, short-range charge fluctuations are quenched, and the typical Umklapp scattering dominated phonon heat transport is recovered. The charge-fluctuations-induced glasslike phonon thermal conductivity implies sizable electron-phonon coupling in $A$V$_3$Sb$_5$.

\end{abstract}
\maketitle
\section{Introduction}
The advent of kagome metals $A$V$_3$Sb$_5$ ($A=$ K, Rb, Cs) possessing intertwined charge density wave (CDW) and superconductivity have attracted extensive research interest lately \cite{Ortiz2019,Ortiz2020,Ortiz2021,Yin2021}. The particular kagome lattice formed by V atoms in $A$V$_3$Sb$_5$ gives rise to various peculiar features in the electronic structure near the Fermi level, such as van Hove singularities and Dirac-like bands \cite{Li2021,Nakayama2021,Liu2021,Ortiz2021Cs,Hu_2022}. An unusual charge order likely associated with time-reversal symmetry breaking appears in $A$V$_3$Sb$_5$ below $T_\mathrm{CDW}\sim$ 80-100 K \cite{Jiang2021a,Wang2021CDW,Shumiya2021,Yang2020,Yu2021,Zheng2021_gating,Chendong2021,Zhou2021,Mielke2021,Yu2021b,Gan2021,Mi_2022,Liang2021CDW,Zhao2021a,Chen2021rotonpair,Xu2021,Li2021a}.  Superconductivity, which likely carries unconventional characters \cite{Li2021NodalSA,Liang2021CDW,Chen2021rotonpair}, arises below $T_\mathrm{c}\sim$ 0.9-2.5 K, and coexists with the CDW phase at ambient pressure \cite{Ortiz2019,Ortiz2020,Ortiz2021,Yin2021}. The intertwined charge and superconducting orders compete in an unusual way, leading to a double-dome superconducting phase diagram in the presence of hyrodstatic pressure \cite{Nguyen2022,Yu2021a,Yu2022pressure,Zhang_pressure,Du2021_pressure,Du2022_pressure,Zhu2021_pressure,Chen2021_pressure,ChenPRL_pressure} and chemical doping \cite{Wang2021_pressure,Oey_doping,Yang_dopinG}. Due to the intricate interplay between lattice geometry, electronic correlations and topological band structures, the driving forces of CDW and superconductivity are still elusive.       

In general, two mechanisms including Peierls-like Fermi surface nesting (FSN) and electron-phonon coupling (EPC), can be relevant in the formation of CDW \cite{Zhu2015_CDW}. It is suggested by density functional theory (DFT) calculations that EPC is not strong enough to produce the observed superconducting $T_\mathrm{c}$ in $A$V$_3$Sb$_5$ \cite{Tan2021}. The weak EPC is in line with scanning  tunnelling  microscopy (STM), angle-resolved photoemission spectroscopy (ARPES) and optical spectroscopy studies, which favor the FSN scenario arsing from the van Hove singularities near the $M$ point \cite{Jiang2021a,Liu2021,Kang2022_FSN,Zhou2021_FSN}. The possibility of strong EPC-induced CDW is also ruled out by the absence of Kohn anomalies in the phonon dispersion at the CDW wave vector as evidenced by hard x-ray scattering experiments \cite{Li2021}. Nevertheless, the importance of momentum-dependent EPC cannot be ignored  as suggested by ARPES, neutron scattering and optical spectroscopy experiments \cite{Luo2022,Xie2022_EPC,Uykur2022_EPC}. In addition, CDW amplitude modes have been observed by Raman scattering measurements, pointing to the important role played by EPC \cite{Liu2022_EPC}. Moreover, recent ARPES measurements found sizable EPC, which is sufficient to account for the superconducting $T_\mathrm{c}$ \cite{Zhong2023_EPC}. A few DFT studies suggest that the EPC driven Jahn-Teller-like distortion is responsible for the CDW, in accordance with the pressure-induced suppression of CDW \cite{Wang2021_EPC,Ptok2022_EPC}.

The EPC can produce significant reduction in phonon thermal conductivity ($\kappa_\mathrm{ph}$), especially in metals and doped semiconductors where electronic and lattice heat conduction are comparable \cite{Liao2015_EPC,Xu2019_EPC,Zhou2020_EPC}. In CDW compounds NbSe$_3$ and 1T-TaS$_2$, the effects of EPC are revealed by distinct anomalies in $\kappa_\mathrm{ph}$ at the CDW transitions \cite{Liu2020_EPC,Yang2019_EPC}. In addition, short-range charge fluctuations above $T_\mathrm{CDW}$ strongly scatter phonons via EPC, leading to glasslike $\kappa_\mathrm{ph}$ that increases weakly with increasing temperature, as found in various CDW materials \cite{Kuo2003_CDW,Smontara_CDW,Murata2015_CDW,Gumeniuk2015_CDW}. Charge fluctuations are also nicely revealed by glasslike phonon thermal transport in stripe-ordered cuprates and nickelates \cite{Hess1999_stripe,Yan2003_stripe}. Sizable charge fluctuations above $T_\mathrm{CDW}$ in $A$V$_3$Sb$_5$ have been suggested by x-ray scattering, nuclear magnetic resonance (NMR) measurements \cite{Chen2022_fluctuation,Zheng2022_fluctuation,Subires2022_fluc}.  In this article, we study the impacts of charge fluctuations on phonon thermal transport in $A$V$_3$Sb$_5$. Above $T_\mathrm{CDW}$, glasslike phonon thermal conductivity appears universally in $A$V$_3$Sb$_5$. The glasslike phonon heat transport can be well described by treating local lattice distortions caused by charge fluctuations via EPC as scattering defects in phonon heat conduction. Our findings suggest that charge fluctuations and EPC play important roles in phonon thermal transport of  $A$V$_3$Sb$_5$, and that thermal conductivity is a sensitive probe in detecting charge fluctuations and EPC.   

\section{Experimental Method}
Single crystalline $A$V$_3$Sb$_5$ samples were grown by a self-flux method in quartz ampoules using alumina crucibles \cite{Li2021NodalSA,Gan2021,Mi_2022}. \textcolor{black}{ Details on structural analysis can be found in the Supplemental Material \cite{supplemental}.}  Crystals with typical dimensions of 2 $\times$ 1.5 $\times$ 0.5 mm$^3$ were chosen in the heat capacity, electrical and thermal transport experiments. Optical images of typical $A$V$_3$Sb$_5$ single crystals are shown in Figs. S1-S3 of the Supplemental Material \cite{supplemental}. The heat capacity measurements were performed in a Physical Property Measurement System (PPMS, Quantum Design Dynacool 9 T) using the relaxation method. The sample coupling constant is over 97\% for all heat capacity experiments. \textcolor{black}{Inaccuracies in heat capacity measurements are less than $\pm$2\% and $\pm$4\% below 10 K and 100 K, respectively. The maximum uncertainty reaches $\sim \pm$10\% at 300 K.} Steady state thermal transport experiments were carried out in PPMS using a home-built setup equipped with one heater and two-thermometer. Errors caused by heat losses in thermal conductivity measurements are estimated to be lower than $\pm$2\% below 150 K. Errors in sample size determination are around $\pm$10\%. The electrical conductivity was recorded using the standard four-probe method in PPMS. Both the thermal gradient and electrical current were directed with the crystal $ab$-plane.

\section{Results and Discussion}
\begin{figure*}
\centering
\includegraphics[scale=0.25]{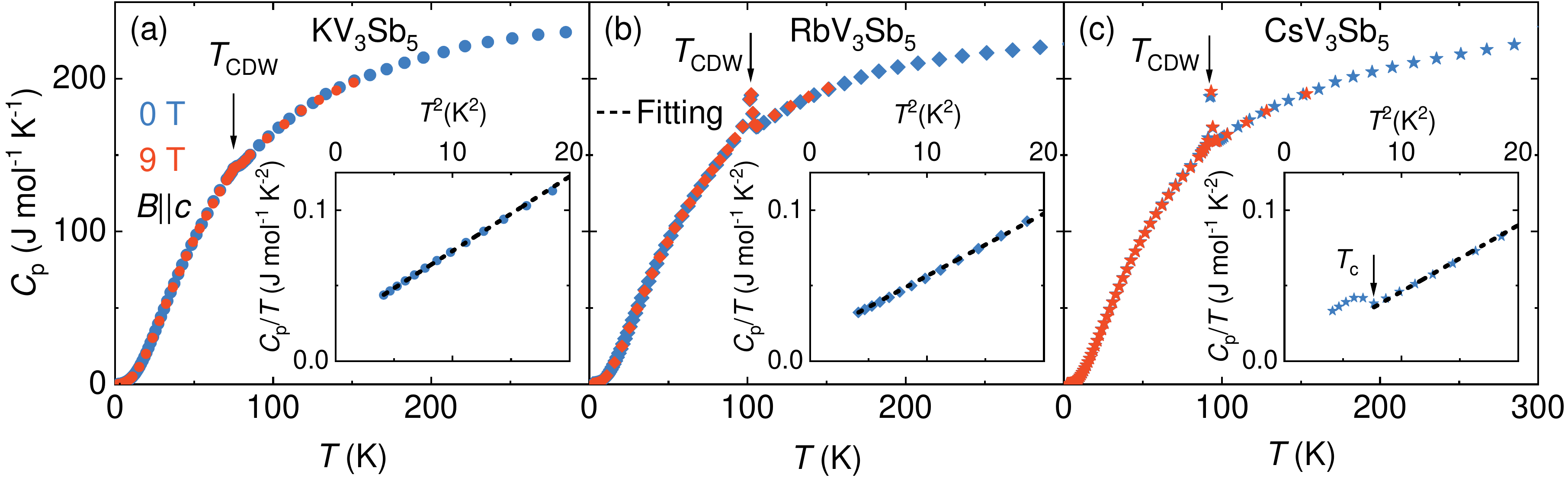}
\caption{Temperature dependence of the specific heat $C_\mathrm{p}$ of (a) KV$_3$Sb$_5$, (b) RbV$_3$Sb$_5$, and (c) CsV$_3$Sb$_5$. The arrows mark out the CDW transitions. \textcolor{black}{ Insets in (a-c) plot the linear fitting (black dash lines) of the specific heat according to $C_\mathrm{p}/T=\gamma+\beta T^2$ within a temperature range of 2-4.5 K.}}
\label{fig1}
\end{figure*}

Figure \ref{fig1} presents the specific heat ($C_\mathrm{p}$) results of typical $A$V$_3$Sb$_5$ samples.  The CDW transitions are evidenced by anomalies in $C_\mathrm{p}$ appearing at $T_\mathrm{CDW}$ = 75, 101, 92 K for KV$_3$Sb$_5$, RbV$_3$Sb$_5$ and CsV$_3$Sb$_5$, respectively, agreeing well with other reports \cite{Ortiz2019,Ortiz2020,Ortiz2021,Yin2021}. The superconducting transition is also seen in CsV$_3$Sb$_5$ at $T_\mathrm{c}=$2.5 K [see the inset of Fig. \ref{fig1}(c)]. Application of an out-of-plane magnetic field of 9 T has negligible impacts on the normal state specific heat, in accordance with the non-magnetic nature of $A$V$_3$Sb$_5$. For non-magnetic metals, the total specific heat ($C_\mathrm{p}$) consists of electronic ($C_\mathrm{e}$) and lattice ($C_\mathrm{ph}$) contributions. At low temperatures, the total specific heat can be described as $C_\mathrm{p}(T)=\gamma T+\beta  T^{3}$ with $C_\mathrm{e}=\gamma T$ and $C_\mathrm{ph}=\beta T^3$. Here, $\gamma$ is the Sommerfeld coefficient and $\beta$ = 12$\pi ^\mathrm{4}NR/5\Theta_\mathrm{D}^{3}$ with $N$, $R$, $\Theta_\mathrm{D}$ representing the number of atoms per unit cell, the ideal gas constant and the Debye temperature, respectively. As shown in the insets of Figs. \ref{fig1}(a-c), the $C_\mathrm{p}/T$ scales linearly with $T^\mathrm{2}$ at low temperatures for all three compounds. The Sommerfeld coefficients and Debye temperatures of $A$V$_3$Sb$_5$ can be extracted accordingly. The obtained results are presented in Table \ref{parameter_of_specific_heat}, agreeing quantitatively with earlier experimental and theoretical studies \cite{Ortiz2021,Yin2021,Tan2021}. Note that the Sommerfeld coefficient of $\sim$ 20 mJ mol$^{-1}$ K$^{-2}$ is relatively large compared to typical metals. The Sommerfeld coefficient is directly related to the density of states (DOS) at the Fermi level \cite{gopal1966specific}. In $A$V$_3$Sb$_5$, van Hove singularities appear near the Fermi level, leading to enhanced DOS \cite{Tan2021}. In addition, DOS can be further enhanced by electron-phonon coupling \cite{tari2003specific}. Another possibility of enhanced Sommerfeld coefficient could be correlation effects associated with kagome lattice. However, electron correlations in $A$V$_3$Sb$_5$ are weak considering the nice agreement between calculated and measured band structures \cite{Zhao2021corelation}. We also note that the Debye temperature does not change systematically among the $A$V$_3$Sb$_5$ series. Normally, the Debye temperature deceases with increasing atomic mass for isostructural compounds. Deviations from this simple expectation are not uncommon \cite{Toher_debye}, and even appear in simple elemental solids \cite{kittel1996-116}. In fact, the Debye temperature is related to the ratio between bulk modulus ($B_S$) and atomic mass ($M$) via $\Theta_\mathrm{D}\propto(B_S/M)^{1/2}$ \cite{Toher_debye}. Non-monotonic variations of $B_S/M$ can thus lead to non-systematic changes in $\Theta_\mathrm{D}$ among isostructural materials.

\begin{table}
\centering
\caption{The Debye temperature ($\Theta_\mathrm{D}$) and Sommerfeld coefficient ($\gamma$) of $A$V$_3$Sb$_5$ extracted from specific heat measurements. 
\label{parameter_of_specific_heat}}
\renewcommand*{\arraystretch}{1.5}
\begin{tabularx}{0.5\textwidth}{
>{\centering\arraybackslash}X 
>{\centering\arraybackslash}X 
>{\centering\arraybackslash}X
>{\centering\arraybackslash}X}
\hline
\hline
Materials & $\Theta_\mathrm{D}$(K)  &$\gamma$(mJ mol$^{-1}$ K$^{-2}$) \\
\hline
KV$_3$Sb$_5$  & 154.81 $\pm$ 1.46 & 23.37 $\pm$ 0.47\\
RbV$_3$Sb$_5$ & 160.32 $\pm$ 1.47 & 15.12 $\pm$ 0.30 \\
CsV$_3$Sb$_5$ & 160.83 $\pm$ 1.48 & 19.30 $\pm$ 0.39 \\
\hline
\hline
\end{tabularx}
\end{table}

\begin{figure*}
\centering
\includegraphics[scale=0.23]{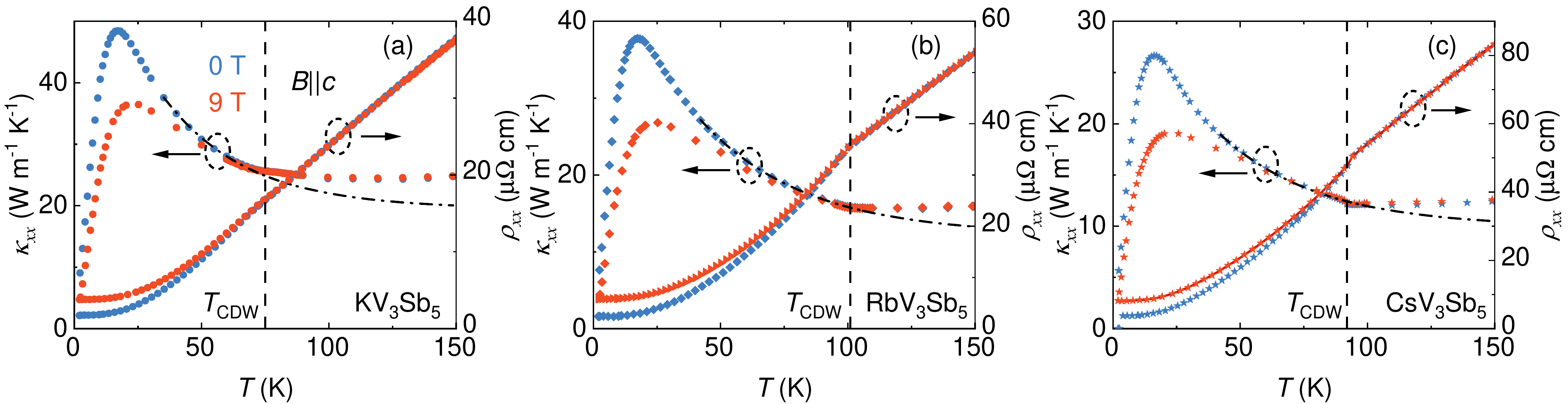}
\caption{In-plane longitudinal thermal conductivity ($\kappa_{xx}$) and electrical resistivity ($\rho_{xx}$) of (a) KV$_3$Sb$_5$, (b) RbV$_3$Sb$_5$,  and (c) CsV$_3$Sb$_5$. The CDW transitions in the three systems are labeled by vertical dash lines. Sizable thermal and electrical magnetoconductivity only appear in the CDW sate. Black dash dot lines are guides to the eye.}
\label{fig2}
\end{figure*}

\begin{figure*}
\centering
\includegraphics[scale=0.25]{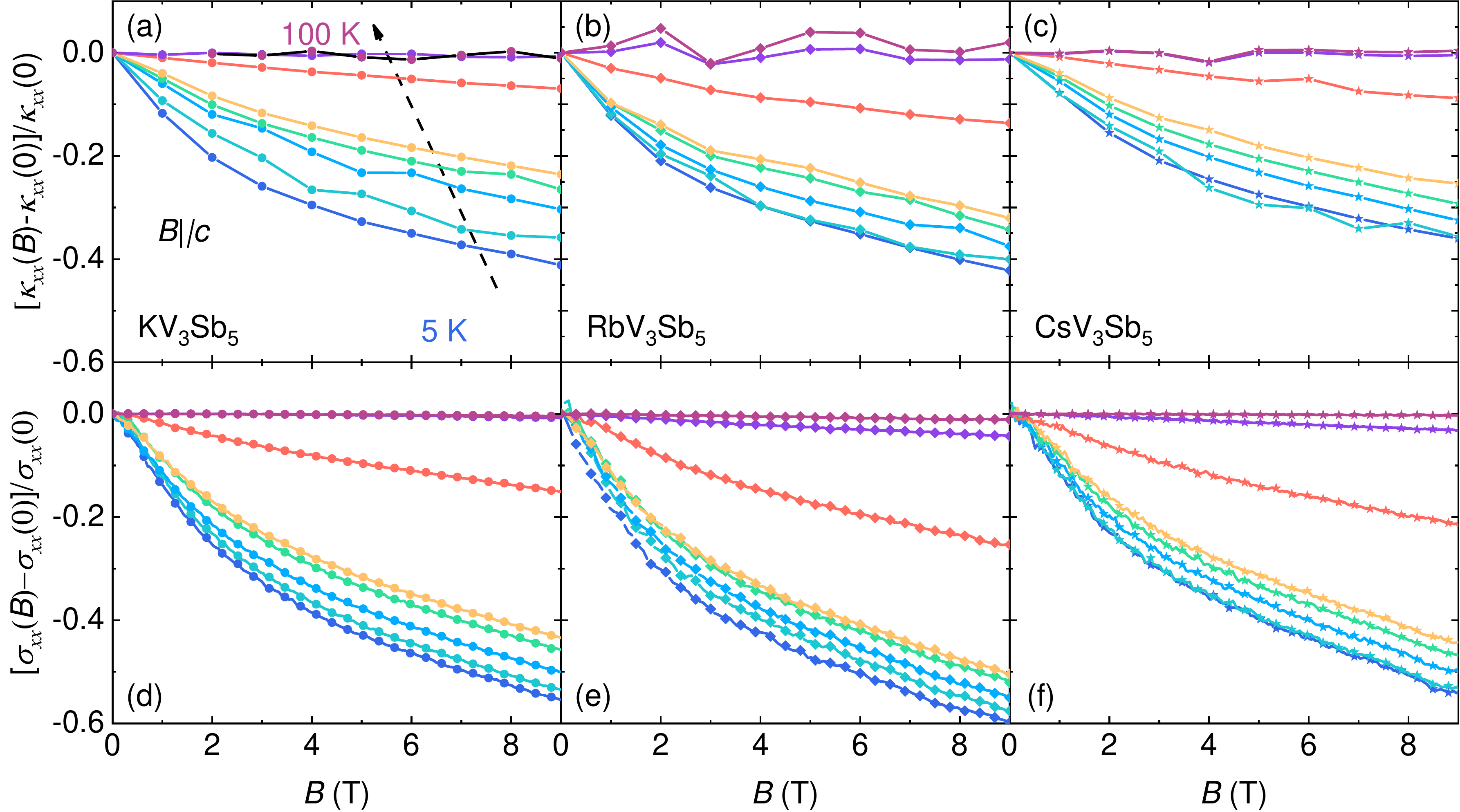}
\caption{Thermal magnetoconductivity (a-c) and electrical magnetoconductivity (d-f) of $A$V$_3$Sb$_5$ measured at fixed temperatures. External magnetic fields were applied along the $c$-axis ($B\parallel c$). In all three compounds, $\kappa_{xx}(B)$ and $\sigma_{xx}(B)$ share similar magnetic field dependence, implying an electronic origin of the thermal magnetoconductivity.  }
\label{fig:3}
\end{figure*}

In Fig. \ref{fig2}, the in-plane longitudinal thermal conductivity ($\kappa_{xx}$) and electrical resistivity ($\rho_{xx}$) of $A$V$_3$Sb$_5$ measured in zero magnetic field and 9 T ($B\parallel c$) are shown. The CDW transitions are manifested by kinks both in $\kappa_{xx}(T)$ and $\rho_{xx}(T)$. Sizable responses to external magnetic fields are only seen in the CDW state both for $\kappa_{xx}$ and $\rho_{xx}$. In fact, the thermal magnetoconductivity is directly linked to electrical magnetoconductivity via the Wiedemann-Franz law. We will analyze thermal and electrical magnetoconductivities in more detail in Fig. \ref{fig:3}.  More interestingly, the temperature dependence of $\kappa_{xx}$ displays distinct behaviors above and below $T_\mathrm{CDW}$ in all three materials. Deep inside the charge ordered state, $\kappa_{xx}$ peaks around 17 K, above which $\kappa_{xx}$ drops quickly in all three compounds. Above $T_\mathrm{CDW}$, $\kappa_{xx}$ deviates from the temperature-dependent trend below $T_\mathrm{CDW}$ (see black dash dot lines in Fig. \ref{fig2}). Right above $T_\mathrm{CDW}$, $\kappa_{xx}$ becomes weakly temperature-dependent, and increases slightly upon further warming. This peculiar temperature dependence of thermal conductivity above $T_\mathrm{CDW}$ is likely originated from charge-fluctuations-induced glasslike phonon thermal transport, as we will discuss in more detail later. Similar temperature-dependent thermal conductivity has also been found in an earlier thermal transport study on CsV$_3$Sb$_5$ \cite{Zhou2021}.

Similar to specific heat, the thermal conductivity of non-magnetic metals mainly contains two parts, i.e., electronic ($\kappa_\mathrm{e}$) and phononic ($\kappa_\mathrm{ph}$) contributions. In metals, the electronic thermal conductivity is directly related to the electrical conductivity ($\sigma$) via the Wiedemann-Franz law $\kappa_\mathrm{e}(T)=L\sigma T$, where $L$ is the Lorenz number. In the Fermi-liquid picture, the Lorenz number is a constant $L_0=\pi^2k_B^2/3e^2=2.44\times10^{-8}$ V$^2$ K$^{-2}$. Although this degenerate limit applies in most metals, substantial deviations of $L$ from $L_0$ may appear at low temperatures or in systems possessing strong correlations \cite{Matusiak2005_LargeL,Muller2008WF,Muller2009WF,Foster2009WF,Crossno2016WF}. Typically, $\kappa_\mathrm{ph}$ is expected to be independent of magnetic field.  The Lorenz number in real materials can then be determined experimentally by comparing the thermal $\kappa(B)$ and electrical $\sigma(B)$ magnetoconductivities measured at fixed temperatures \cite{White1958_lorenz,Armitage1969_lorenz,Uher1974_lorenz,Sharp2001_lorenz,Lukas2012_lorenz},

\begin{equation}
    \kappa_{xx}(B)=LT\sigma_{xx}(B)+\kappa_\mathrm{ph}.
    \label{eq:2}
\end{equation}
Here, we focus on the longitudinal terms assuming that $L$ is independent of magnetic field. If Eq. \ref{eq:2} is valid, $\kappa_{xx}(B)$ scales linearly with $\sigma_{xx}(B)T$ with the slop being the Lorenz number.  

\begin{figure*}
\centering
\includegraphics[scale=0.25]{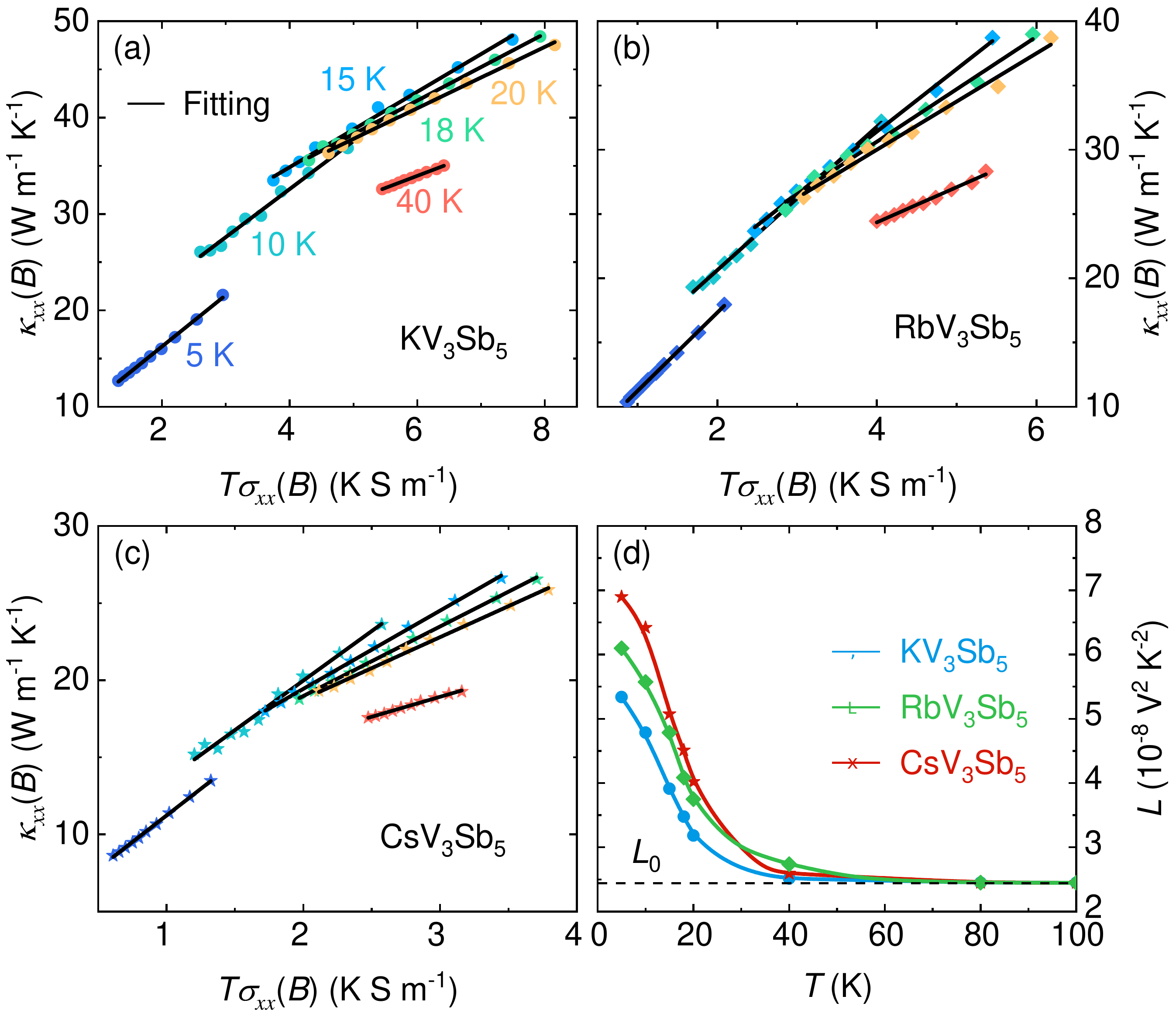}
\caption{(a-c) Scaling of $\kappa_{xx}(B)$ vs $\sigma_{xx}(B)T$ using the magnetoconductivity data shown in Fig. \ref{fig:3}. Solid black lines are linear fittings, from which the Lorenz number can be extracted according to Eq. \ref{eq:2}.  (d) The estimated Lorenz number as a function of temperature. Solid lines are guide to the eye. Below about 80 K, the Lorenz number deviates from the degenerate value $L_0$ and is strongly enhanced at low temperatures.}
\label{fig:4}
\end{figure*}

In Fig. \ref{fig:3}, the longitudinal thermal magnetoconductivity $\kappa_{xx}(B)$ and electrical magnetoconductivity $\sigma_{xx}(B)$ of $A$V$_3$Sb$_5$ recorded at selective temperatures are presented. At high temperatures above 80 K, both $\kappa_{xx}(B)$ and $\sigma_{xx}(B)$ show negligible magnetic field dependence for all three members in $A$V$_3$Sb$_5$. Substantial thermal and electrical magnetoconductivities develop at lower temperatures. Note that although $\kappa_{xx}(B)$ and $\sigma_{xx}(B)$ differ in magnitudes, they share similar functional forms, suggesting the possibility to extract the Lorenz number from Eq. \ref{eq:2}. Indeed, as shown in Figs. \ref{fig:4}(a-c), $\kappa_{xx}(B)$ and $\sigma_{xx}(B)T$ follow the linear relationship of Eq. \ref{eq:2} at all temperatures measured here. By linear fitting of $\kappa_{xx}(B)$ vs $\sigma_{xx}(B)T$, the estimated temperature-dependent Lorenz number is shown in Fig. \ref{fig:4}(d). It is seen that above 80 K, $L$ stays closely to the degenerate limit $L_0$. Below 80 K, $L$ deviates from $L_0$ and grows continuously with cooling. Rather large values of  $L$ above 5 V$^2$ K$^{-2}$ are found at low temperatures in all three materials. Large enhancements of $L$ have  been found in the pseudogap state of high-$T_\mathrm{c}$ cuprate and iron-based superconductors, likely originating from the depletion of density of states (DOS) at the Fermi level  \cite{Minami2003_LargeL,Matusiak2005_LargeL,Matusiak2015_LargeL}. Similarly, reduced DOS caused by the momentum-dependent CDW gaps near the Fermi energy can lead to enhanced $L$ in the charge ordered state of $A$V$_3$Sb$_5$. Notably, in strongly interacting systems, transport behaviors can enter the hydrodynamic region, in which electrical and thermal transports are decoupled, leading to giant enhancements of $L$ \cite{Muller2008WF,Muller2009WF,Foster2009WF}. This highly unusual phenomenon has been observed in graphene, in which the quasi-relativistic fermions near the charge-neutrality point form a strongly coupled Dirac fluid that can be described by hydrodynamics \cite{Crossno2016WF}. In $A$V$_3$Sb$_5$, the connections between nontrivial topological properties and substantially enhanced Lorenz number in the CDW state need further investigations. 

\begin{figure}
\centering
\includegraphics[scale=0.3]{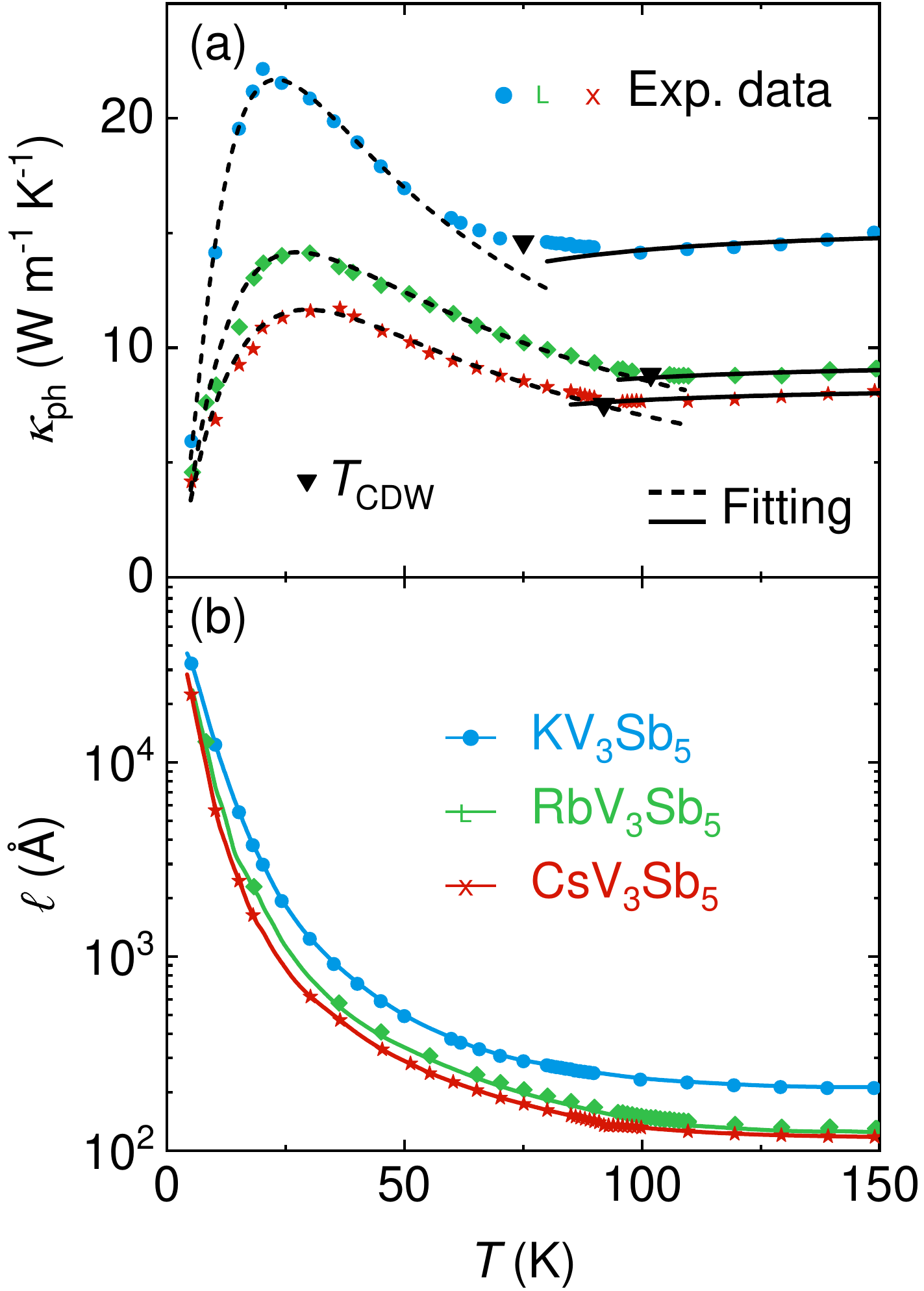}
\caption{(a) Temperature dependence of phonon thermal conductivity $\kappa_\mathrm{ph}$ in $A$V$_3$Sb$_5$ obtained after subtracting the electronic contributions from the total thermal conductivity. The black triangles mark the CDW transitions.  Black dash lines are theoretical modelings of $\kappa_\mathrm{ph}$ below $T_\mathrm{CDW}$ following the Debye-Callaway approach by considering the conventional phonon scattering mechanisms. Above $T_\mathrm{CDW}$, $\kappa_\mathrm{ph}$ is well described by incorporating the scattering of phonons by charge-fluctuations-induced local lattice distortions via EPC (solid black lines).  (b)Temperature dependence of the phonon mean free path ($\ell$) of $A$V$_3$Sb$_5$.} 
\label{fig:5}
\end{figure}

\begin{table*}
\centering
\caption{Fitting parameters of $\kappa_\mathrm{ph}$ in the Debye-Callaway model by considering different phonon scattering mechanisms above and below the CDW transition.
\label{tab:best-fitting parameters}}
\renewcommand*{\arraystretch}{1.5}

\begin{tabularx}{1\textwidth}{
>{\centering\arraybackslash}p{5em}
>{\centering\arraybackslash}p{5em} 
>{\centering\arraybackslash}p{5em}
>{\centering\arraybackslash}p{0.2em}
>{\centering\arraybackslash}p{5em}
>{\centering\arraybackslash}p{2em}
>{\centering\arraybackslash}p{6em}
>{\centering\arraybackslash}p{4em}
>{\centering\arraybackslash}p{0.2em}
>{\centering\arraybackslash}p{5em}
>{\centering\arraybackslash}p{4em}
>{\centering\arraybackslash}p{4em}
>{\centering\arraybackslash}p{4em}}
\hline
\hline
\multirow{2}{*}{Materials} & \multirow{2}{*}{$v_s$ (m s$^{-1}$)} & \multirow{2}{*}{$d$ (10$^{-6}$m)} & \multirow{2}{*}{}& \multicolumn{4}{c}{$T<T_\mathrm{CDW}$} & \multirow{2}{*}{}& \multicolumn{4}{c}{$T>T_\mathrm{CDW}$}\\
\cline{5-8}
\cline{10-13}
& & & &$A(10^{-42}\mathrm{s}^{3})$ & $n$ & $B(10^{-18}\mathrm{K}^{-1}\mathrm{s})$ & $b$ & & $A(10^{-16}\mathrm{s})$ & $n$ & $B$ & $b$\\ 
\hline
KV$_3$Sb$_5$  & 1511.79(5) & 4.15(1) & & 3.63(5) & 4 & 6.15(1) & 4.25(1) & & 5.68(5) & 2 & 0 & 0\\
RbV$_3$Sb$_5$ & 1612.81(8) & 4.39(1) & & 8.09(1) & 4 & 5.19(1) & 3.65(1) & & 9.40(1) & 2 & 0 & 0\\
CsV$_3$Sb$_5$ & 1595.99(8) & 4.45(1) & & 12.43(1) & 4 & 6.55(1) & 2.45(1) & & 10.51(2) & 2 & 0 & 0\\
\hline
\hline
\end{tabularx}
\end{table*}

With the Lorenz number obtained, the phonon thermal conductivity $\kappa_\mathrm{ph}$ can be evaluated accordingly using $\kappa_\mathrm{ph}=\kappa_{xx}-L\sigma_{xx} T$. Here $\kappa_{xx}$ is the total thermal conductivity shown in Fig. \ref{fig2}. The temperature dependence of the Lorenz number $L(T)$ below $T_\mathrm{CDW}$ was estimated by simple interpolations of the data shown in Fig. \ref{fig:4}(d). Above $T_\mathrm{CDW}$, $L=L_0$ was used in the calculation. The results of $\kappa_\mathrm{ph}$ are displayed in Fig. \ref{fig:5}(a). Despite of the metallic nature of $A$V$_3$Sb$_5$, $\kappa_\mathrm{ph}$ contributes considerablely to the total thermal conductivity.   The most notable feature in $\kappa_\mathrm{ph}$ is its contrasting temperature dependence above and below $T_\mathrm{CDW}$, suggesting different phonon scattering mechanisms. In the kinetic phonon gas picture, the phonon thermal conductivity of a three dimensional system is linked to the phonon specific heat via $\kappa_\mathrm{ph}=\frac{1}{3}C_\mathrm{ph}v_s\ell$, where $v_s$ and $\ell$ are the average sound velocity and the phonon mean free path, respectively.   Within the Debye scenario of phonon specific heat, the phonon heat conduction can be captured well by the Debye-Callaway model \cite{Callaway1959}:
\begin{equation}
    \kappa_\mathrm{ph}(T) = \frac{k_B^4}{2\pi^2 v_s\hbar^3} T^3 \int_0^{\Theta_\mathrm{D}/T} \frac{x^4 e^x}{(e^x-1)^2} \tau(\omega,T)dx,
    \label{Debye-calloway model}
\end{equation}
 where $x=\hbar\omega/k_BT$. The scattering events of phonons, such as Umklapp scattering in the three phonon processes, point-defect scattering and boundary scattering, are taking into account in the relaxation time $\tau(\omega,T)=\ell/v_s$. The total relaxation time can be considered as the Matthiessen sum of all possible scattering mechanisms assuming that different scattering events contribute independently:

\begin{equation}
\begin{split}
    \tau(\omega,T)^{-1}&=\tau^{-1}_b+\tau^{-1}_{pd}+\tau^{-1}_U,\\ 
    \tau^{-1}_b=v_sd^{-1}, \tau^{-1}_{pd} &=A\omega^n, \tau^{-1}_U=B\omega^2Te^{-\Theta_\mathrm{D}/bT},
    \label{eq:4}
\end{split}
\end{equation}
where $\tau_b$, $\tau_{pd}$ and $\tau_U$ are relaxation times of boundary, point-defect and Umklapp scatterings, repsectively. And $d$ represents the grain size, $A$, $n$, $B$ and $b$ are numerical constants associated with corresponding scattering mechanisms. At low temperatures, the phonon mean free path is limited by sample boundary, which is weakly temperature-dependent. The $\kappa_\mathrm{ph}$ is then proportional to $C_\mathrm{ph}$, and increases with temperature as $\kappa_\mathrm{ph}(T)\sim T^3$. At high temperatures, Umklapp scattering dominates and $\kappa_\mathrm{ph}(T)$ decreases approximately as $T^{-1}$ with increasing temperature. A characteristic peak is typically found in $\kappa_\mathrm{ph}(T)$  at the intermediate temperature region when the dominating scattering mechanism switches roles. This is the case for $A$V$_3$Sb$_5$ in the charge ordered state [see Fig. \ref{fig:5}(a)]. Indeed, by considering the usual scattering processes shown in Eq. \ref{eq:4},  the Debye-Callaway approach well describes $\kappa_\mathrm{ph}(T)$ of $A$V$_3$Sb$_5$  below $T_\mathrm{CDW}$ [see the black dash lines in Fig. \ref{fig:5}(a)]. The fitting parameters are shown in Table \ref{tab:best-fitting parameters}. The Debye temperature was taken from the analysis of specific heat as displayed in Table \ref{parameter_of_specific_heat}. And the average sound velocity was estimated from the Debye temperatures as $v_s=\Theta_\mathrm{D}(\frac{k_B}{\hbar})(6\pi^2N/V)^{-1/3}$, where $N$ and $V$ are number of atoms and volume per unit cell. For the point-defect scattering rate, the form of $\tau_{pd}^{-1}=A\omega^4$ ($n=4$ in Eq. \ref{eq:4}) was used, assuming that the typical sizes of lattice point defects are smaller than phonon wavelength.       

Above $T_\mathrm{CDW}$, things become different and $\kappa_\mathrm{ph}(T)$ increases smoothly with increasing temperature. Such temperature dependence of $\kappa_\mathrm{ph}(T)$ is reminiscent of glasslike phonon thermal transport when $\ell$ approaches the inter-atomic distances \cite{Anderson1972,Allen1999,Mukhopadhyay2018}.  Although typically found in disordered materials, glasslike phonon thermal transport can arise in single crystals possessing strong charge/orbital or spin fluctuations, which effectively lead to 'disordered' lattice via electron-phonon or spin-phonon couplings \cite{Kuo2003_CDW,Smontara_CDW,Murata2015_CDW,Gumeniuk2015_CDW,Sharam2004_spin,Li2013_spin}. Substantial charge fluctuations above $T_\mathrm{CDW}$ have been experimentally suggested in (Cs,Rb)V$_3$Sb$_5$ by X-ray scattering, NMR measurements \cite{Chen2022_fluctuation,Zheng2022_fluctuation,Subires2022_fluc,Luo2022_fluctuation,Song2022_fluctuation}. The correlation length ($\xi$) of short-range charge fluctuations extends up to about 5 unit cells $\xi\sim$ 30 $\mathrm{\mathring{A}}$ \cite{Subires2022_fluc}. In Fig. \ref{fig:5}(b), we estimated the phonon mean free path by $\ell=3\kappa_\mathrm{ph}/C_\mathrm{ph}v_s$. Note that this estimation only provides a rough guidance to an order of magnitude for $\ell$. Above $T_\mathrm{CDW}$, the phonon mean free path ($\ell\sim$ 100 $\mathrm{\mathring{A}}$) is weakly temperature-dependent. Importantly, $\xi$ is comparable to $\ell$. The short-range charge fluctuations can thus serve as effective scattering centers of phonons. Short-range charge orders, either static or dynamic, can induce local lattice distortions via EPC, leading to strong phonon damping and glasslike phonon thermal conductivity \cite{Hess1999_stripe}.  The short-range charge-fluctuations-induced local lattice distortions can be viewed as lattice defects, which enter the point-defect scattering rate as $\tau_{pd}^{-1}=A\omega^2$ ($n=2$ in Eq. \ref{eq:4}) if the defect sizes approach the phonon wavelength \cite{Joshi1979,Murias2011}. Using this form of $\tau_{pd}^{-1}$ and ignoring Umklapp scattering, the glasslike $\kappa_\mathrm{ph}(T)$ above  $T_\mathrm{CDW}$ can be well reproduced in the Debye-Callaway approach [see solid black lines in Fig. \ref{fig:5}(a)], implying the dominant role played by charge fluctuations in phonon scattering events. Details of the fitting parameters are listed in Table \ref{tab:best-fitting parameters}.       

The peculiar charge-fluctuations-induced glasslike phonon heat transport indicates sizable EPC in $A$V$_3$Sb$_5$. Although the exact strength of EPC is still elusive, it appears that EPC plays important roles in $A$V$_3$Sb$_5$ \cite{Luo2022,Xie2022_EPC,Uykur2022_EPC,Zhong2023_EPC,Wang2021_EPC,Ptok2022_EPC}. Recent thermal diffuse scattering and inelastic X-ray scattering experiments suggest that the CDW transition is an order-disorder type, which could explain the absence of the Kohn anomaly, and is likely originated from strong EPC or non-adiabatic Peierls instability in the strong coupling limit \cite{Subires2022_fluc}. The order-disorder scenario is in line with the thermal conductivity results, which show glasslike transport with damped phonon propagations in the disordered state above $T_\mathrm{CDW}$. Below $T_\mathrm{CDW}$, establishment of the long-range charge order produces correlation lengths much larger than the phonon mean free path, allowing the recovery of typical phonon thermal conductivity. As seen in Fig. \ref{fig:5}(b), $\ell$ increases rapidly in the long-range CDW state, leading to the usual Umklapp scattering dominated phonon heat transport. 

\section{Conclusions}
We have studied the thermal conductivity of the kagome metals $A$V$_3$Sb$_5$ in detail. By linear scaling of thermal and electrical magnetoconductivities, large enhancements of the Lorenz number are found in the CDW state. From the experimentally determined Lorenz number, the phonon thermal conductivity was successfully extracted.  Unlike the usual phonon heat transport, the phonon thermal conductivity shows glasslike behavior above the CDW transition, which appears to be generic in all three members in $A$V$_3$Sb$_5$. The glasslike phonon thermal transport is very likely caused by charge fluctuations and sizable electron-phonon coupling. \textcolor{black}{Further detailed theoretical calculations from first principles are desired to quantitatively extract the strength of electron-phonon coupling and the impacts of electron-phonon interaction on the phonon heat transport of $A$V$_3$Sb$_5$. Together with thermal transport, Raman, ARPES and x-ray scattering experiments, the role played by electron-phonon coupling in the formation of CDW and superconductivity in $A$V$_3$Sb$_5$ could be singled out. }

\section{Acknowledgments}
 We thank Guiwen Wang and Yan Liu at the Analytical and Testing Center of Chongqing University for technical support. This work has been supported by National Natural Science Foundation of China (Grant Nos. 11904040, 52125103, 52071041, 12004254, 12004056, 11674384, 11974065), Chinesisch-Deutsche Mobilit\"atsprogamm of Chinesisch-Deutsche Zentrum f\"ur Wissenschaftsf\"orderung (Grant No. M-0496), Chongqing Research Program of Basic Research and Frontier Technology, China (Grant No. cstc2020jcyj-msxmX0263). Y. Guo acknowledges the support by the Major Research Plan of the National Natural Science Foundation of China (No. 92065201).
 
\bibliographystyle{apsrev4-2}

\end{document}